# Trade-offs in the design and communication of flood-risk information

**Authors:** Courtney M. Cooper[1*], Sanjib Sharma[1], Robert E. Nicholas[1,2], Klaus Keller[1,3,4]

[1]Earth and Environmental Systems Institute, The Pennsylvania State University, University Park, PA, USA
[2]Department of Meteorology and Atmospheric Science, The Pennsylvania State University, University Park, PA, USA
[3]Department of Geosciences, The Pennsylvania State University, University Park, PA, USA
[4]Thayer School of Engineering, Dartmouth College, Hanover, NH, USA
* Corresponding author

## Highlights

1. We identify ten desirable features of flood-risk estimates designed to guide decision-making.
2. The available products providing flood-risk information have shortcomings that can result in poor decisions.
3. Comprehensive uncertainty characterization is a critical next step to improve the credibility of flood-risk information.
4. Co-production is a promising approach to improve the relevance of flood-risk information.

## Abstract

There is an increasingly urgent need to develop knowledge and practices to manage climate risks. For example, flood-risk information can inform household decisions such as purchasing a home or flood insurance. However, flood-risk estimates are deeply uncertain, meaning that they are subject to sizeable disagreement. Available flood-risk estimates provide inconsistent and incomplete information and pose communication challenges. The effects of different choices of design and communication options can create confusion in decision-making processes. The climate services literature includes insights into desirable features for producing information that is credible and relevant. Using examples of riverine (fluvial) flood-risk information products and studies in the United States, we assess how existing risk characterizations integrate desirable features outlined in the climate services literature. Improved characterization and communication of decision-relevant (and often deep) uncertainties, including those arising from human decisions, is a crucial next step. We argue that producing relevant flood-risk information requires applying principles of open science and co-production.

## Keywords

Flooding is the most deadly and costly of natural disasters leading to deaths and severe economic consequences[1]. Improved characterization of flood risk can help decision-makers to better manage and prepare for future flooding[2,3]. Although sophisticated approaches to flood-risk characterization are available, many challenges limit their credibility and relevance in decision-making[4–6]. First, integrating projections of urbanization and climate change is difficult[7,8]. Second, flood hazards stem from diverse and interacting mechanisms (pluvial, fluvial, or coastal) spanning a range of spatial and temporal scales[8–10]. Third, inconsistencies exist in the representations of the primary elements of flood risk—hazard, exposure, and vulnerability[11]. Hazard refers to the extent and depth of a flood[9,12]. Exposure includes the people and properties at risk, and vulnerability the characteristics of the people and property at risk[13]. Fourth, many interrelated factors such as regional climate patterns influence model performance[9].

These challenges impact flood-risk estimates. For one, estimates are deeply uncertain, meaning "the system model and input parameters to the system model are not known or widely agreed on by the stakeholders to the decision"[14]. In addition, market considerations often undermine scientific objectives. For example, products based on proprietary code and data are not reproducible[15–17]. The regulatory timeframes for updating design standards also often lag scientific and technological advancements[18,19]. Many open questions remain about translating outputs from flood-risk characterizations and models into flood-risk products (henceforth 'products') to inform decision-making[5,6,20].

These barriers and challenges are not unique to flooding. Similar challenges are discussed across the climate services literature[6,21,22]. Information providers (henceforth 'providers') and information users (henceforth 'users') often have different needs, priorities, and values[6]. For example, the prevailing norms of science may lead providers to focus on the technical and scientific aspects of design rather than designs to improve decisions[6]. Previous research suggests that users are more likely to use the information they perceive as accurate, credible, salient, and timely[5,20,23–26]. Yet, there is little guidance available about what features are needed to produce usable information. The potential consequences of flooding and the role that information can play in informing flood protection decisions warrant critical evaluation of information usability.

Recently, information design has, in principle, shifted from hazard- to risk-focused approaches[27,28]. Risk-based approaches inform decisions such as household flood protection decisions (e.g., purchasing flood insurance[29,30]). A property's flood risk depends on many factors, including the topographic setting, property characteristics, flood defenses in the watershed, and floodplain inundation dynamics[31]. In the United States (US), several providers characterize household flood risk. Federal standards, set by the Federal Emergency Management Agency (FEMA), guide the design of Flood Insurance Rate Maps (FIRMs), which inform flood-risk protection policies and regulations[32,33]. FIRMs do not integrate future change and only account for riverine flooding and flooding associated with storm surges[34]. The First Street Foundation's Flood Factor® incorporates some climate change information and estimates flood hazards from multiple sources of flooding[35,36]. Estimates provided by academic and commercial settings apply yet another range of design choices[4,9].

We provide insight and direction for designing information for assessing household flood risk. We focus on the requirements of producing more usable and comprehensive fluvial (riverine) flood-risk information, noting that modeling multiple sources of flooding and compound flooding adds complexity. We begin by reviewing the literature on climate services to identify desirable information design and communication features. We then compare available product descriptions and recent examples in the US, including FIRMs, Flood Factor®, and academic research (see Table 1). The considered academic research examples, while not developed as decision-making products, demonstrate variations in design objectives.

**Results**

      **Information with high spatial resolution and wide coverage.** Decisions about property-level flood protection require information at that same spatial scale. Flood hazards are sensitive to local conditions, and therefore challenging to accurately estimate using large-scale models[27]. Detailed characterizations of property-level risk require studies for individual properties including an assessment of characteristics such as the elevation, age, and construction style of the specific property[37]. Assessment of individual property characteristics across large geographic areas (e.g., nations) through individual studies is costly and resource intensive, if not impossible[31]. Large-scale models can be updated more frequently at lower cost, relative to approaches that require detailed studies. Still, these estimates can be prone to errors arising from local conditions that influence an individual property's flood risk [27,38].

    Although scientific questions remain, technical advances allow models to resolve flood-risk estimates for individual properties across large areas[7,9,31]. Local data availability is often a limiting factor in scaling up these studies[39]. How best to assess and validate models for specific decision-problems is the subject of fast-moving and ongoing research[3,12,29,40]. Maximizing household scale information availability for an entire country requires accurate yet cost-effective approaches[31]. Local studies remain critical for assessing the performance of national flood-risk products. For instance, providers can assess the adequacy and accuracy of downscaled climate information at regional scales and then apply insights from these studies at larger scales[41].

      **Open science principles.** According to the National Academies of Sciences, Engineering, and Medicine, open science ensures "the free availability and usability of scholarly publications, the data that result from scholarly research, and the methodologies, including code or algorithms, that were used to generate those data"[42]. Institutions have developed approaches for promoting open science. For instance, the Findability, Accessibility, Interpretability, and Reuse (FAIR) Framework outlines principles of open science[43]. At the same time, criticisms of open science focus on its multiple meanings and the differences between principles and practice[15]. For information to be open access in a democratic process, the methods and methodologies used to gain the knowledge must be transparent[44].

    Accessible information relies on intuitive, well-designed, and up-to-date websites with easy-to-follow links to additional resources[45,46]. Flood Factor® has received attention for the product's accessibility because it is easily accessible via a stand-alone web interface and through the real estate websites such as Realtor.com. On the Flood Factor® website, users can compare their property risk designated by FEMA with their Flood Factor® risk score[35]. The product provides information about historic flooding and future risk. It includes resources about how users can mitigate flood risk, such as a cost calculator. The output from traditional academic studies is often difficult to interpret or unavailable.

    Data, methods, computer codes, analysis plans, conflicts of interest, and value judgments determine a product's transparency[44]. Transparency is desirable for comparison and modification[26]. High levels of transparency can increase the product's long-term usability and improve communication[47,48]. However, factors such as reluctance to share code and data for proprietary reasons constrain transparency[24,44]. Transparency between users and providers is complex. When provided with technical details about the design, users may become confused or lose trust in the product[44]. We focused on whether traceable code and data allow providers to reproduce a product.

Simplicity is desirable across climate services[49,50]. Simpler product designs are easier to reproduce and modify. However, studies using a single model structure can fail to capture uncertainty in design choices, such as the choice of governing equations and numerical transformations[51]. Complex models often demand computational resources, larger input datasets, and a high degree of model parameterizations[9]. Trade-offs occur when balancing between simple and complex design choices, especially as technical advancements allow products to capture dynamics about changing climate and environmental conditions[12].

**Integrating landscape change, climate change, vulnerability, and exposure.** Comprehensive risk information integrates landscape change, climate change, vulnerability, and exposure. Including these features leads to more complex frameworks, however, excluding them can result in less accurate information[9,12]. These features are essential for capturing complex dynamics between land, human societies, and water[8]. For instance, land disturbance associated with urban development and upstream channel erosion can increase sediment in the channel and diminish flood conveyance capacity, overshadowing climate, and land-use effects[52].

**Uncertainty characterization remains important but fragmented.** Uncertainty is a key property of flood risk estimates[52]. There are (sometimes heated) arguments on how to characterize and communicate this uncertainty[29,52]. Some scholars argue that the technical risk communication literature tends to treat uncertainty as an epistemological gap that can and should be reduced to zero[53]. Others argue that uncertainty is a launching point for new discourses and approaches to understanding risk[54]. Uncertainty characterization requires assessing a range of uncertainty sources.

Uncertainties can arise, for example, from choice of model structures, model parameters, and unresolved processes related to human decisions and biophysical processes[9,55–57]. One goal of uncertainty characterization is to determine the most relevant uncertainties for a particular decision problem[58]. For example, Zarekarizi *et al.*[29] shows how interactions between economic, engineering, and earth systems drive the total uncertainty in the cost and benefits of elevating a house. Importantly, neglecting uncertainty can produce downward bias in risk estimates[51,59,60]. Uncertainty characterization provides a way to better understanding information quality[61].

Key epistemic uncertainty sources include model structure, model parameter, channel geometry, surface topography, and human decisions. Aleatoric uncertainty arises from natural variability (e.g., the random nature of thunderstorms that drive extreme precipitation)[62]. For flood-hazard estimates, existing research suggests that, although spatially variable[52], discharge, channel parameter, and topography are the most dominant sources of uncertainty[63]. However, other research highlights the dominant influence of climate model-driven uncertainty in flood inundation projections[10,64]. Because providers struggle to characterize the cascade of uncertainty sources[9], less is known about the sources and effects of uncertainty and their propagation and interactions in flood-risk estimates.

Neglecting deep uncertainties can underestimate flood hazards and risks[29,65]. Deep uncertainty stems, for example, from model limitations such as uncertain land-use changes. Flood protection and adaptation measures illustrate the importance of uncertainty characterization. First Street Foundation integrates local information about flood adaptation and mitigation measures, such as the locations of flood defense structures within Flood Factor®. However, gaps in this knowledge exist because Flood Factor® does not yet account for whether these measures will perform to their designed level under future climate conditions. Uncertainty characterization methods can help to improve understanding of deep uncertainty[14].

**Legally accepted information for decision-making.** Legally accepted products are used to establish standards such as insurance mandates. For example, property owners with federally backed mortgages in high-risk flood zones as designated by FEMA must have flood insurance[30]. We consider the product's used to make regulatory decisions legally accepted. While legally accepted, FIRMs do not include many of the desirable features. FEMA's risk rating structure has changed minimally since the NFIP's inception in 1968[66]. The changes primarily coincide with technological advances in mapping conventions, for instance, FIRM digitization[67].

Shortcomings in FIRMs, including a need to improve flood-risk information, are well-documented[68–71]. The US Association of Floodplain Managers estimates that FEMA's flood mapping cost $10.6 billion (in 2019-dollar values) between 1969 and 2020. Yet, the program only covers 33% of the rivers and streams at variable levels of quality, and many existing maps are outdated[37]. Further, the legal flood mapping standards often lag behind the latest science[18]. To improve the usability of FIRMs in decision-making, some states have developed more user-friendly approaches for exploring FIRMs and related information. For instance, North Carolina's Floodplain Mapping Program maintains a web-based decision support tool where users can view property-scale risk information such as the flood zone and estimated cost of flooding impacts[72]. Although limited to the information provided by FEMA, this platform offers easy access to information about flood insurance and mitigation.

Efforts to improve FIRMs are ongoing and FEMA offers several alternative products, such as information offered through RiskMAP, to provide additional information about flooding[73]. The NFIP estimates a property's risk rating by combining hazard estimates derived from FIRMs with individual property characteristics. FEMA began implementing a new rating system, Risk Rating 2.0. Once fully implemented, Risk Rating 2.0 will be the first major design overhaul to the NFIP's risk rating structure[66]. The overarching goals are to reflect actuarial rates by leveraging industry best practices, being more equitable, and making rates easier to understand[66]. However, the proposed framework relies on proprietary models that might not align with open science principles or meet users' needs.

**Discussion**

Producing flood-risk information for decision-making requires making many design and communication choices. These choices include navigating synergies and trade-offs in product objectives, such as those between complexity and transparency. The considered examples navigate tradeoffs in design and communication choices differently (Table 2), reflecting the deep uncertainties in flood-risk characterization. Before producing larger and more complex models and frameworks, it is important to understand the decision-relevant uncertainties.

FIRMs include detailed local engineering studies but lack spatial coverage in many areas and are not sensitive to changing future conditions. Although Flood Factor® improves upon FIRMs, it potentially ignores local characteristics that remain difficult to capture through large models. Across the examples, we found low levels of transparency, highly complex frameworks, and limited focus on uncertainty characterization. Many examples include climate change projections, but do not integrate changing landscape conditions, such as the life expectancies of flood defense structures. Continued improvements in the capacity for cross-product comparison, a greater emphasis on uncertainty characterization, and collaborative web-based product design platforms can help to further advance open science principles.

**Towards more comprehensive flood-risk product design.** Fig. 1 illustrates design characteristics for two of the examples (FEMA and First Street Foundation products) and an idealized product based on a more comprehensive product design. The figure highlights sources of potentially decision-relevant uncertainties and the system components needed to estimate fluvial flood risk. Regardless of the methods and methodologies, the primary components and key uncertainties identified in Fig. 1 are critical to producing credible flood-risk products. Reproducing and comparing products and studies is a complex, if not infeasible, undertaking[48,74]. The low comparability between products can prevent users from understanding the full extent of the risk they face. Recent analyses comparing Flood Factor® risk scores with risk estimates produced by FEMA show dramatic increases in risk based on Flood Factor® projections relative to FEMA risk estimates[7,36]. Improving product comparability could make it easier to understand their strengths and weaknesses.

The examples we assessed consider a limit subset of uncertainties. Many studies assess the impacts of uncertainty in flood peaks[39] but not the interactions between uncertainty sources and their propagation into flood risk estimates[55,56]. Or, often they ignore the effects of uncertainty in exposure and vulnerability[75,76]. As a result, it can be difficult to understand the impacts of factors such as model boundary conditions (e.g., landscape features and river geometry) and model structural uncertainties. Uncertainties might be partially resolved, for example, by analyzing the occurrence of hazards, the growth patterns of exposed people and assets, and the effects of actions taken to reduce risk[38].

Model structural uncertainty and parametric uncertainty can influence all linkages displayed in Fig. 1. Several studies outline the potential of characterizing model structural uncertainty through multi-model systems[51,77]. For instance, Bayesian model averaging offers a framework to integrate information based on the credibility of each model output[56]. Providers can characterize parametric uncertainties through model calibration[78,79] and use case studies to evaluate interactions[38,39,80]. Although characterizing parametric uncertainties is complex, it can provide insights into decision-relevant uncertainties, with the potential to lead to more straightforward and transparent designs. Primary avenues to improving uncertainty characterization include identifying decision-relevant uncertainties, understanding how uncertainty propagates, and quantifying the contribution of individual uncertainty sources and their associated interactions[81].

Finally, collaborative web-based platforms can reduce barriers to comparing risk estimates and uncertainty characterization. Across climate science domains, enlisting and maintaining critical and informed providers and users is essential for meeting open science standards[42]. Funders can make it easier to produce open source products by providing web services for sharing information like those envisioned by the Open Water Data Initiative[82], an initiative that aims to build a community around water resources data[83]. Another platform, called SWATShare, allows users to upload and share models, run simulations, and visualize results[84]. Providers need consistent methods for describing product designs and outputs[85,86]. Shared online platforms can enable collaboration and decrease barriers to sharing code and data.

**Communicating flood-risk information.** Assessing how communication styles impact decision-making requires careful research as even small differences in visual styles can lead to different decisions[6]. Yarnal *et al.*[46] reveals that that color schemes and map keys of existing climate information products were challenging to understand. A survey of flood-risk management experts found a dislike for communication in "x-year" terms. Instead, most thought the public would respond better to descriptive words such as "high risk" or "moderate risk"[88]. In another study, information presented in a map format decreased study participants' ability to accurately select houses with lower hazards relative to the same information presented in graphs or tables[89].

Communication styles varied across the considered examples. FIRMs illustrate flood hazards through binary flood zone maps, while the Flood Factor® style is point based and shows descriptive risk scores. Fig. 2 recreates three different visual styles to illustrate the range of styles. Fig. 2a displays flood hazard in a style similar to FIRMs, Fig. 2b aligns with an academic study[90] that renders the flood hazard information provided by FIRMs as continuous rather than binary. In Fig. 2c, which aligns with the visual style of Flood Factor®, the hazard description does not correspond with the other two panels. The underlying product design differs, and it assigns a score to specific structures rather than assessing risk for the property. Notably, risk to both the property and structures are decision relevant. The influence that different visual styles have on decision-making is unknown.

The house labeled "1" falls within the 500-year flood zone in Fig. 2a, which is in the visual style of FIRMs. However, in Fig. 2b, where the same hazard information is communicated as a continuous flood probability, and House 1 appears to be in a higher flood probability zone. Flood zones communicate that a home has "at least" the specified level of hazard (e.g., at least a 0.2% chance of flooding in any given year). Continuous flood hazard probability maps give a more precise estimate of the likelihood of flooding[90]. In Fig. 2c, House 1 has an "extreme risk" and draws from different underlying information than Fig. 2a&b. The different information presented in each panel could influence how users perceive risk.

Other communication questions arise from Fig. 2. The classification scheme in Fig. 2c may unintentionally provide a false sense of certainty for property owners with properties outside designated flood zones[32,68,69]. For instance, although House 2 is in a hazard zone across the three panels, flood-risk could be under-estimated because of the possibility of losing access to critical infrastructure such as hospitals if floodwaters covered the road. The visual style in Fig. 2c may not clearly communicate the risk associated with road flooding. Fully characterizing risk requires tools that demonstrate indirect risk, such as threats to infrastructure.

Decision support systems are increasingly capable of presenting detailed risk information through user-friendly interfaces[30]. These tools require careful attention to visual formats[46,89]. The First Street Foundation is the first organization, to our knowledge, to implement a consistent, nationwide option for visualizing household flood risk. Flood Factor® draws attention to climate change and uses communication best practices to help users interpret information. In the US, this information may be particularly valuable to prospective home buyers as many areas of the country do not have mandatory flood disclosure policies[91]. Important questions remain about how Flood Factor® will influence decision-making. Although First Street Foundation's disclaimers outline potential risk and allowable uses[92], it remains unclear how the product will influence property values.

**Co-produced flood-risk products.** When produced absent of dialogue, information is often not applicable to decision-making and can lead to mistrust between users and providers[18,30,87]. Co-produced information is less likely to generate costly legal responses and more likely to create long-term community benefits[68,93]. In its best form, co-production can shift communication from best guess estimates to explicit acceptance of deep uncertainty[94]. Co-production can lead to products designed as thinking aids intended to build knowledge about adaptation and mitigation goals[95].

Questions remain about how principles of co-production can be applied to flood-risk characterization. Co-production requires building long-term trusting relationships, as such it often occurs through small, case study applications[96]. We did not identify an example of information co-produced with providers and users. First Street Foundation relies on some level of co-production as they partner with academic institutions. Updating FIRMs requires a process that includes opportunities for public comment and protest[97]. Recently, examples of collaborative flood-risk modeling have emerged at regional and local scales[18,98].

Leveraging collaborative online environments could lead to more informed providers and users. In the US, FEMA plays an important leadership role in developing co-production initiatives. In the future, FEMA could expand their web-based tools using an approach like Flood Factor®. Available flood maps could be offered in multiple formats that allow users to explore the costs and benefits of different flood-protection measures[45].

We offer a starting point for producing more comprehensive flood-risk information products. The desirable features will continue to evolve with scientific and technological advances. Moving forward, attention on open science principles, co-production, and uncertainty characterization can lead to more comprehensive products. These improved products can help users to better understand their flood risk. Improved information design alone is unlikely to change decision-making processes, attention to process is critical.

**Methods**

To evaluate tradeoffs in objectives across the examples, we reviewed the growing literature about climate services. We identified features associated with high information quality for decision-making[e.g.,4,6,24]. We narrowed these features to a list of primary desirable features needed to produce credible and relevant flood-risk products (described in Supplementary Materials). We consider a feature desirable when—in principle—the feature could help support providers to design more credible products and users to make more informed decisions. After identifying these features, we assessed how the reviewed examples integrate the feature. We focused our evaluation on information features that can be assessed without conducting primary research. For example, understanding whether users access information is critical, however, assessing usability requires long-term evaluation of users' knowledge and preferences[99]. The features identified here are non-exhaustive but representative of the vast literature.

Figure 1 summarizes the primary sources of uncertainty and key components needed. We developed the figure by reviewing related research. Product designs utilize a range of methods and methodologies including different models of land surface hydrology (ranging from lumped to physically distributed) and river hydraulics (from one-dimensional to two-dimensional). Regardless of the selected method, the key uncertainties and components remain the same. This figure could serve as a starting point for developing future flood-risk information. The components and uncertainty sources are described in the Supplemental Materials.

**Acknowledgments**

This work was co-supported by the Penn State Initiative for Resilient Communities (PSIRC) through a Strategic Plan seed grant from the Penn State Office of the Provost, the Center for Climate Risk Management (CLIMA), the Rock Ethics Institute, Penn State Law, the Hamer Center for Community Design, the National Oceanic and Atmospheric Administration, Climate Program Office under grant NA16OAR4310179, and the MultiSector Dynamics program areas of the U.S. Department of Energy, Office of Science, Office of Biological and Environmental Research as part of the multi-program, collaborative Integrated Coastal Modeling (ICoM) project. Any conclusions or recommendations expressed in this material are those of the authors and do not necessarily reflect the views of the funding entities. We thank Lisa Domenica lulo, Lara Fowler, and Casey Helgeson for their inputs. Katerina Kostadinova helped with figure design. Any errors and opinions are, of course, those of the authors.

**Author Contribution Statement**

Investigation, writing—original draft preparation; C.M.C., resources, supervision, project administration, and funding acquisition; K.K. Conceptualization, methodology, formal analysis, visualization, writing—review and editing; C.M.C., S.S., R.E.N, and K.K. All authors have read and agreed to the published version of the manuscript.

# Figures and Tables

## Table 1. Example products and studies

| Example overview | Primary limitations |
|---|---|
| FEMA Flood Insurance Rate Map (FIRM) | |
| -Flood hazard maps based on local engineering studies<br>-Legally accepted in decision-making<br>-Basis for floodplain management, mitigation, and insurance activities of the National Flood Insurance Program (NFIP) | -Does not integrate risks associated with future climate change<br>-Distill riverine flood hazards to a single loss mechanism based on historic observations<br>-Do not characterize uncertainty<br>-Assessing changing flood hazards through a patchwork of local studies is time consuming and expensive |
| The First Street Foundation Flood Factor® | |
| -Characterizes risk through for individual properties across the continental US and integrates an approximation of projected climate change effects<br>-Estimates risk for households across the continental US<br>-Output is easily accessed online<br>-Accounts for flooding from fluvial, pluvial, and coastal sources<br>-Incorporates some sources of climate uncertainty<br>-Provides information about historic flooding and projects future risk<br>-Described in peer-reviewed journals and reviewed by expert panel | -Underlying code and data are not open source<br>-Silent on the effects of potential key uncertainties, such as topographic uncertainties that may manifest in the data and methods<br>-Some properties may be incorrectly labeled<br>-Uses an adaptation database, historical claims, and flood reports to assess flood hazard and risk estimates, although the exact approach is unclear because the underlying code and data are not provided |
| Wobus et al. (2019) | |
| -Project change in expected annual fluvial flood damages across the continental US<br>-Estimates damages for the entire distribution of flood events rather than specific flood zones<br>-Provides a robust understanding of exposure at a large spatial scale<br>-Integrate different flood scenarios<br>-Includes the effects of adaptation measures | -Inherits limitations from the underlying FEMA-based hazard estimates |
| Rajib et al. (2020) | |
| Use hydrologic modeling to generate streamflow simulations, hydrodynamic modeling to estimate | -Does not integrate climate change<br>-Limited spatial coverage |
| water surface elevations, and terrain analysis to map flood inundation extents<br>-Relative to others improves the integration of processes<br>-Future work can factor in other global datasets<br>-Produced with the principles of FAIR in mind<br>-Relatively simple design | |
| Judi et al. (2018) | |
| -Integrate downscaled climate scenarios, hydrology, and flood consequence to estimate flood risk<br>-Provides local-scale, actionable information under changing climate conditions<br>-Transparent discussion about neglected uncertainty sources<br>-Output provided at a scale that is meaningful for household decision-making<br>-Integrates landscape characteristics | -Applied at a small spatial scale<br>-Dependent on local datasets |
| Zarekarizi et al. (2021) | |
| -Interpolate flood probabilities across elevations to produce continuous flood-probability maps based on FEMA hazard estimates.<br>-Relatively simple to implement<br>-Publicly accessible open-source code<br>-Continuous flood-probability maps<br>-Visual to communicate different flooding probabilities inside the 100- or 500- year floodplains delineated by FIRMs | -Inherits limitations from the underlying FEMA-based hazard estimates |
| Liu et al. (2015) | |
| -Apply a "bottom-up" case study approach to estimate flood damages from extreme events in the midwestern US<br>-Accounts for differences in local conditions<br>-Assessment of the sensitivities of flood damages to different economic and climate change outlooks | -Scaling to other locations is difficult because sufficient data is often unavailable<br>-Relies on historical data records to uncover a statistical relationship between disaster loss and individual flood risk factors |

**Table 2.** Desirable features as identified in select flood-risk information products.

| Desired Features | FIRMs | Flood Factor | Wobus et al. 2019 | Judi et al. 2018 | Rajib et al. 2020 | Zarekarizi et al. 2021 | Liu et al. 2015 |
|---|---|---|---|---|---|---|---|
| **Legally Accepted*** Product is legally binding | ■■■ | □ | □ | □ | □ | □ | □ |
| **Transparency** Open source data and code | □ | □ | □ | □ | □ | ▨ | ▨ |
| **Access** Accessible at no cost | ■■■ | ■■■ | ▨ | ▨ | □ | ▨ | ▨ |
| **Resolution** Resolved to the property level | ▨ | ■■■ | ▨ | ▨ | ▨ | ▨ | ▨ |
| **Coverage** Available across the CONUS | ▨ | ■■■ | ▨ | □ | ▨ | □ | □ |
| **Climate Change** Integrates climate change projections | □ | ■■■ | ■■■ | ■■■ | □ | □ | ■■■ |
| **Vulnerability and Exposure** Considers vulnerability and exposure | ▨ | ▨ | ■■■ | ■■■ | ▨ | □ | ■■■ |
| **Landscape Change** Integrates land-cover/use changes | □ | ▨ | ▨ | ▨ | ▨ | □ | ▨ |
| **Simplicity** Simple to implement | ▨ | □ | ▨ | □ | ▨ | ▨ | ▨ |
| **Uncertainty Characterization** Characterizes two or more sources | □ | ▨ | ▨ | ▨ | ▨ | □ | ▨ |

*Feature is binary

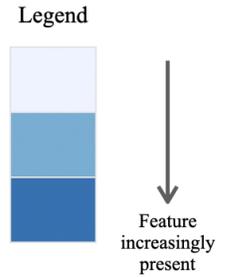

Legend: Feature increasingly present

*Note.* A full description of the feature criteria is provided in the supplementary materials.

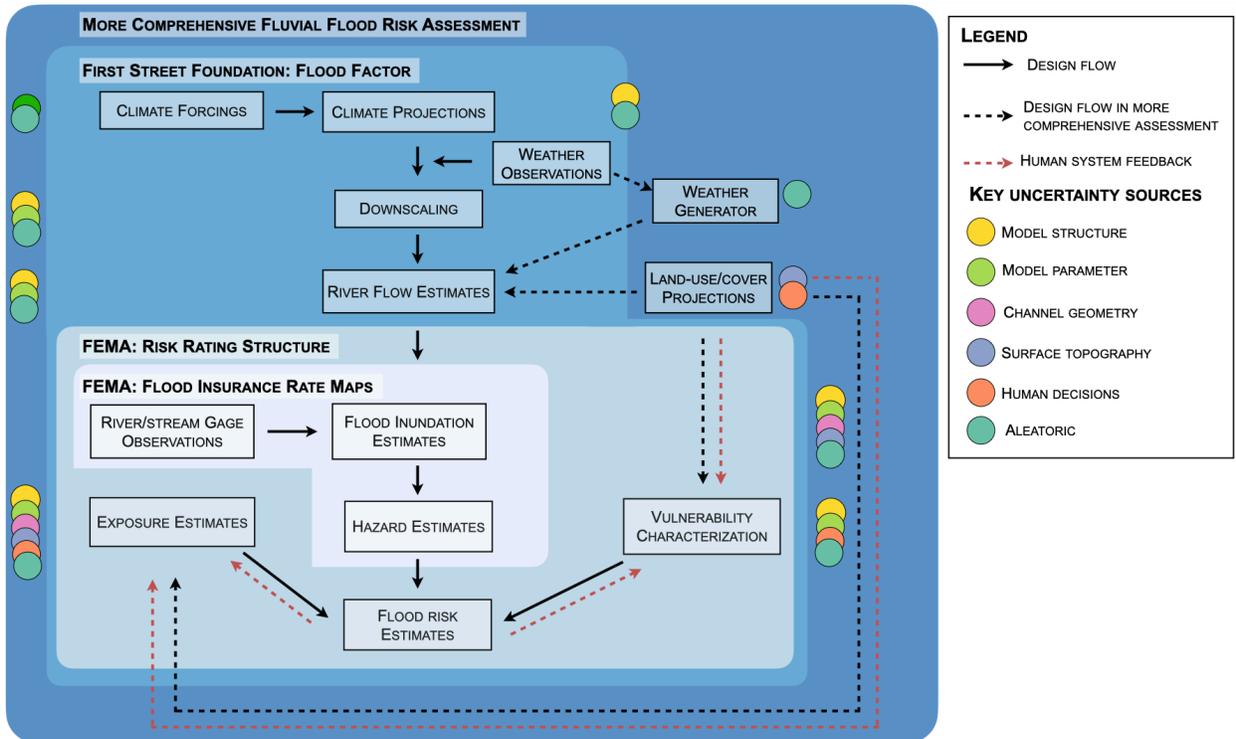

**Figure 1.** Flow diagram illustrating key decision-relevant uncertainties for different flood-risk products. The darkest blue box includes the features of an idealized and relatively comprehensive approach to understanding riverine flood risk. A full description of the figure is provided in the Supplementary Materials.

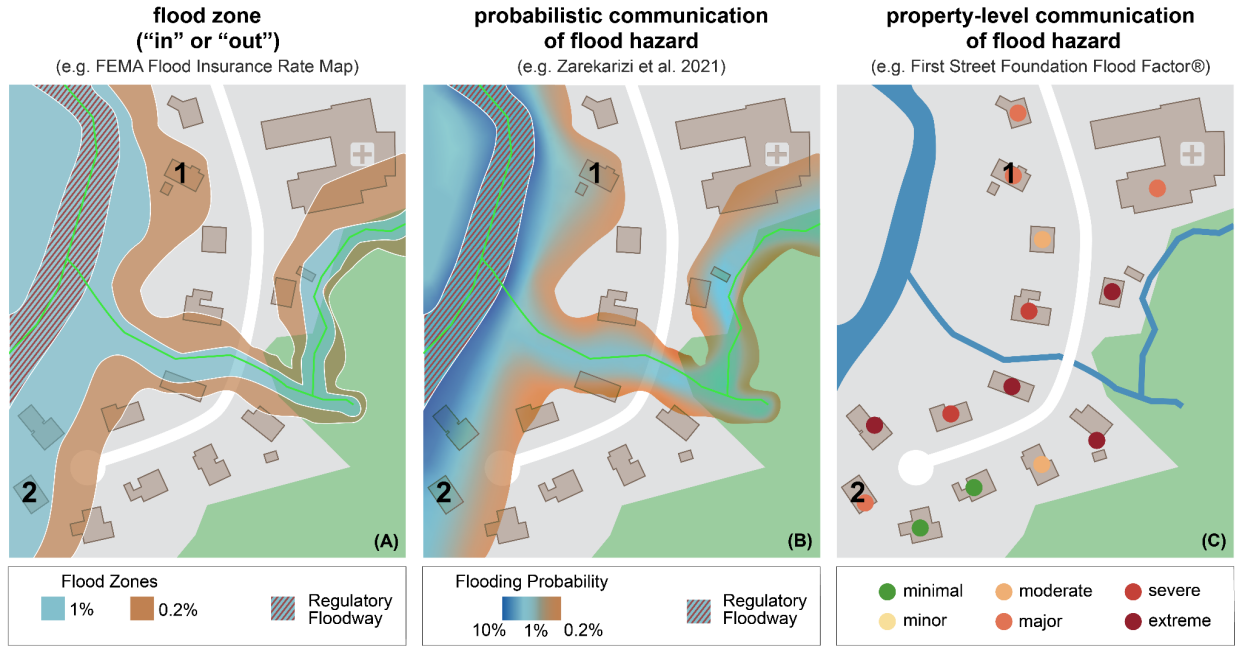

**Figure 2.** Illustration of the different visual styles used to communicate flood risk. Panel A is similar in style to FEMA Flood Insurance Rate Maps; Panel B is based on an academic study that communicates the information in FIRMs as continuous rather than binary flood risk; Panel C is similar in style to the First Street Foundation and uses risk labels rather than flood zones and integrates information about climate sensitivity. The three different styles can lead users to draw different conclusions about household flood risk.
Note. Panels include example citations

**Supplemental Materials**

**Criteria description for Table 1.**

| Feature | Criteria |
|---|---|
| **Legal Acceptability** <br> Product is legally binding | Not legally binding <br> — <br> Legal binding |
| **Transparency** <br> Open source data and code | No data and code <br> Clear data and code accessibility statement, <br> Open source code and data |
| **Access** <br> Accessible at no cost | Only accessible through proprietary sources <br> Open access <br> Available online at no cost |
| **Resolution** <br> Resolved to the property level | Not available at property level <br> Available at property level but not fully-resolved <br> Resolved to the property level |
| **Coverage** <br> Available across the CONUS | Available only for a case study <br> Available for several states or a region of the CONUS <br> Available across the CONUS |
| **Climate Change** <br> Integrates climate change projections | Does not integrate climate change projections <br> Integrates a single climate projection <br> Integrates multiple climate projections |
| **Vulnerability and Exposure** <br> Considers vulnerability and exposure | Does not consider vulnerability or exposure <br> Considers either vulnerability or exposure <br> Considers vulnerability and exposure |
| **Landscape Change** <br> Integrates land-cover/use changes | No land-cover/use <br> Land-cover/use but no change <br> Integrates land-cover/use changes |
| **Simplicity** <br> Simple to implement | Complex model/framework <br> Complexity in some aspects <br> Relatively simple to implement |
| **Uncertainty Characterization** <br> Characterizes two or more sources | No uncertainty characterization <br> Characterizes one source of uncertainty <br> Characterizes two or more sources of uncertainty |

**Description of Figure 1.**

1. **Climate**

Several studies highlight the dominant impact of climate model-driven uncertainty in flood peak estimates[1,2]. Often, flood-risk products integrate only a subset of potentially relevant uncertainties. For instance, the Flood Factor product samples a relatively small subset of GCM outputs under the RCP 4.5 scenarios[3]. Previous studies have shown that considering a small subset of climate projections can drastically undersample the deep uncertainties[4,5]. GCMs do not resolve fine-scale hydrometeorological processes (i.e., local weather), particularly precipitation extremes[6,7]. Nonetheless, GCM output provides valuable information to estimate future flood risk under climate change.

Modelers often use dynamical or statistical downscaling methods to generate climate projections for local-scale analyses[8]. The choice of downscaling method (statistical v. dynamical) has implications for the kinds of uncertainty that propagate from climate model outputs to land surface hydrologic and river hydraulic models. Combined, differences in downscaling techniques and model resolutions can contribute to sizable variations in climate forcing (e.g., extreme rainfall) in river flow models. Choosing a subset of these variations undersamples the deep uncertainties that arise from the range of possible approaches.

Weather generators provide potential avenues for producing probabilistic inferences about future flooding scenarios instead of static estimates of future flooding[9]. Weather generators include high levels of aleatoric uncertainty because they cannot fully resolve all the weather and climate phenomena relevant to flood-producing rainfall events. Downscaling methods and weather generators have limited capabilities to generate high-resolution, precise estimates of global change, but they are necessary for locally relevant flood hazard projections[9,10]. Despite the limitations, information producers often use these methods when designing flood-risk products.

2. **Land surface hydrologic and river hydraulic modeling**

Hydrological models are used to estimate river flow within a catchment by representing complex hydrological cycle dynamics using various parameters and sets of mathematical equations. River flow estimates provide boundary conditions to run the river hydraulic model and generate estimates of flood inundation depth and extent. In traditional approaches to estimating river flow and flood inundation, information producers manually adjust a subset of model parameters to calibrate models[11,12]. This approximation may fail to identify the decision-relevant parameters and can drastically undersample parametric uncertainty. Surrogate methods (e.g., Gaussian process-based emulators) focus on addressing parametric uncertainty[12]. However, building a process-based emulator for a high-dimensional model is challenging[13]. Recent research efforts on Bayesian Statistical Inference implemented by stochastic algorithms such as a fast sequential Markov Chain Monte Carlo (MCMC) provide a probabilistic framework for characterizing parametric uncertainties[14]. Bayesian calibration of hydrodynamic models can be computationally expensive, particularly when estimates for a large number of model parameters are needed.

3. **Uncertainty sources in river flow and flood inundation characterizations**

Inputs for river flow projections include information such as temperature and precipitation. River flow projections provide the boundary condition for the river hydraulics model to estimate flood inundation extent and depth inundation models[11,15]. Limitations in observational records create challenges when developing a land surface hydrologic and river hydraulic model for any catchment[16]. Extreme floods, the floods with the most severe socioeconomic consequences, are associated with limited data records and require modeling with sparsely distributed river monitoring infrastructure[17]. Additional challenges arise when using empirical relationships to characterize conditions in ungauged catchments[2]. Modeling approaches have evolved to include methods for capturing antecedent moisture conditions and model regional processes such as snowmelt[11,18].

4. **Topographic and river geometry data**

Flood inundation estimates require reliable surface topographic data and geometric representation of the river channel[19–21]. Digital elevation models (DEMs) provide detailed topographic data. The quality and resolution of DEMs affect the accuracy of the extracted topographic features[20]. High-resolution DEMs that fully represent the topographic features at local relevant scales are often unavailable. Scientists produce high-resolution DEMs by applying remote sensing technology such as Light Detection and Ranging (LiDAR)[22]. However, LiDAR cannot penetrate the water surface to yield bathymetric results[19]. Hence, LiDAR data do not capture submerged river channel features. LiDAR and DEM information are often integrated with field surveyed bathymetry data to improve the representation of riverbed topography. Despite improved methods, bathymetric uncertainty increases with water depth and river turbulence[21].

5. **Bathymetric and topographic uncertainty**

Uncertainty in information about bathymetric conditions impacts flood inundation estimates. Topographic uncertainty impacts both land-use/cover characterizations and flood inundation estimates. One reason for these uncertainties is that bathymetric and topographic data are not collected uniformly for integration in flood-hazard characterizations[20]. Further, the desired resolution and scale of these datasets are spatially variable.

National scale data about land-use/cover are available through the National Land Cover Database (NLCD,[23,24]), but must be integrated with river flow and flood inundation models[25]. The NLCD is updated every five years and provides nationwide data based on 30-meter resolution Landsat data. Land cover projections datasets are also available but at coarser spatial resolutions[26]. Recent studies have used historical land cover datasets to create land-use change scenarios to project historical development trends into the future[18].

6. **Vulnerability and exposure**

Vulnerability and exposure are highly dynamic features of flood risk[27], with local policies, governmental decisions, and the initiative of individuals, such as the implementation of structural mitigation measures[28] or land-use change and urbanization[29], affecting vulnerability and exposure over time. In turn, these changes can alter the frequency and magnitude of flooding[27]. Channelization projects and levees can increase community vulnerability by increasing urbanization and floodplain development[18]. If the flood control/mitigation structure fails, increased development (exposure) may amplify losses. The reliability of flood-mitigation infrastructure under future climatic conditions is unclear. Further, because human decisions impact the climate system, climate information that does not integrate the uncertainties of human decision-making may lead to overconfident estimates of flood risk[30].

7. **Human decision uncertainty**

Methods and modeling capabilities have limited capacity to account for feedback from dynamic elements of the human system. Available methods for assessing vulnerability, exposure, and uncertainties in human decisions often rely on qualitative methods, historical insurance claims [31–33], or national scale survey data such as the US Census[34–36]. Information producers aggregate data from national surveys to spatial scales that limit the applicability of the data for local decision problems. Further, they do not provide insights into future change. Interactions between human decisions and flood hazards are difficult to capture in models. Although necessary for understanding risk, estimates of exposure and vulnerability are subject to false assumptions or bias, which threatens the external validity of any product[27]. Uncertainty characterization alone is insufficient for communicating the multi-dimensional complexities of flood risk.